\documentclass[10pt,final,twocolumn]{IEEEtran}

\ifCLASSINFOpdf
\else
\fi

\usepackage{graphicx}
\usepackage{subfigure}
\usepackage{amsmath}
\usepackage{amsthm}
\usepackage{dsfont}
\usepackage{algorithmic}
\usepackage{xspace}
\usepackage[ruled,vlined]{algorithm2e}
\usepackage{color}
\usepackage[font=small]{caption}
\usepackage{mathrsfs}

\usepackage{times}
\usepackage{psfrag}
\usepackage{amsfonts,amssymb,amsmath}
\usepackage[innercaption]{sidecap}
\usepackage{graphicx}
\usepackage{bm}
\usepackage{url}

\def\Blue#1{\textcolor{black}{#1}}

\def\er{\mathrm{e}}
\def\dr{\mathrm{d}}
\def\MSE{\mathrm{MSE}}

\def\sr{\mathrm{s}}

\def\xb{\mathbf{x}}

\def\zb{\mathbf{z}}
\def\hb{\mathbf{h}}

\def\kb{\mathbf{k}}

\def\bb{\mathbf{b}}

\def\nb{\mathbf{n}}

\def\Xb{\mathbf{X}}
\def\Ab{\mathbf{A}}

\def\Bb{\mathbf{B}}

\def\Db{\mathbf{D}}

\def\Rds{\mathds{R}}
\def\Eds{\mathds{E}}

\def\betab{\boldsymbol{\beta}}

\def\alphab{\boldsymbol{\alpha}}

\newtheorem{definitionPan}{\textbf{Definition}}

\begin{document}

\title{Wireless for Control: Over-the-Air Controller}

\author{Pangun Park, Piergiuseppe Di Marco, and Carlo Fischione \thanks{P. Park is with the Department of Radio and Information Communications Engineering, Chungnam National University, Korea (e-mail: pgpark@cnu.ac.kr). P. Di Marco is with the Department of Information Engineering, Computer Science and Mathematics, University of L'Aquila, Italy (e-mail: piergiuseppe.dimarco@univaq.it). C. Fischione is with the Department of Network and Systems Engineering, School of Engineering, KTH Royal Institute of Technology, Sweden (e-mail: carlofi@kth.se).}}

\IEEEcompsoctitleabstractindextext{

\begin{abstract}
\Blue{In closed-loop wireless control systems, the state-of-the-art approach prescribes that a controller receives by wireless communications the individual sensor measurements, and then sends the computed control signal to the actuators. We propose an over-the-air controller scheme where all sensors attached to the plant simultaneously transmit scaled sensing signals directly to the actuator; then the feedback control signal is computed partially over the air and partially by a scaling operation at the actuator. Such over-the-air controller essentially adopts the over-the-air computation concept to compute the control signal for closed-loop wireless control systems. In contrast to the state-of-the-art sensor-to-controller and controller-to-actuator communication approach, the over-the-air controller exploits the superposition properties of multiple-access wireless channels to complete the communication and computation of a large number of sensing signals in a single communication resource unit. Therefore, the proposed scheme can obtain significant benefits in terms of low actuation delay and low wireless resource utilization by a simple network architecture that does not require a dedicated controller. Numerical results show that our proposed over-the-air controller achieves a huge widening of the stability region in terms of sampling time and delay, and a significant reduction of the computation error of the control signal.}
\end{abstract}




}

\maketitle


\IEEEdisplaynotcompsoctitleabstractindextext
\IEEEpeerreviewmaketitle

\section{Introduction}
\Blue{Time-critical operations over a wireless network are the heart of essential infrastructures for monitoring and control of systems for factory automation, process control, and power distribution~\cite{Park18, Lunze14}}. In Wireless Networked Control Systems (WNCS), distributed sensors, controllers, and actuators exchange sensing and actuating signals over a wireless network to achieve a control objective~\cite{Park18}. The WNCSs are fundamentally different from traditional distributed systems since the network dynamics, such as time-varying capacity, node faults, and stochastic delay and reliability, significantly affect the physical dynamics of the control system.

In the state-of-the-art WNCS architecture, a number of sensor nodes sample the plant states and then each sensor sends the samples to the controller; when all sensor data arrive, the controller calculates the control signal and transmit it to actuators; actuators operate the received control signal to manipulate the plant~\cite{Lunze14}. {The delay and losses of sensors-to-controller and controller-to-actuators links are crucial for the stability of the system and may degrade the control performance.} \Blue{Furthermore, as advanced systems using microsensors and embedded computers determine an increased density and scale of sensor networks,  more bandwidth is required to collect the sensor measurements.} Since the design of WNCS requires the simultaneous interaction of communication, computation, and control aspects, in addition to physical phenomena, new computation and communication architectures and protocols are needed for closed-loop wireless control systems~\cite{Park18}.

Within the communication community, the over-the-air computation paradigm has been recently proposed and investigated to efficiently compute linear functions of sensor measurements by utilizing the superposition property of multiple-access channels~\cite{Abari16, Henrik20}. Opening an innovative field of applications for this paradigm, we recognize that a linear feedback controller, which is one of the most practical controller design approaches, computes its control signal in the form of a weighted sum of plant state measurements~\cite{Goodwin00}. 

In this paper, we propose the concept of an over-the-air controller (AirCont) by adopting the over-the-air computation concept to compute the control signal of the closed-loop wireless control systems. Since AirCont uses the multiple-access channel to compute the weighted sum of the sensor data, it fundamentally overcomes the sensor-to-controller and controller-to-actuator state-of-the-art architecture, by introducing a fundamentally new direct sensor-to-actuator architecture. We first analyze the achievable benefits of the proposed AirCont scheme in terms of control stability with the operational feasibility constraints. We also investigate the computation error of existing power control approaches to compute the control signal subject to the peak power constraints of nodes. We demonstrate that AirCont yields significant improvements in terms of the achievable stability region and reduces the computation error for different power limits, channel noise, and a number of nodes compared to the state-of-the-art. Finally, we illustrate the use of AirCont in a case study for the control of a ball and beam system.

The rest of the paper is organized as follows. In Section~\ref{sec:system}, we describe our control system model and the controller design schemes over a wireless network model. Following that, we analyze the control stability region with the feasible operating constraints and the computation error using different schemes in Section~\ref{sec:analysis}. Then, we evaluate the performance benefits using AirCont compared to the state-of-the-art control schemes in Section~\ref{sec:eval}. Finally, we present conclusions and discussion.

\emph{Notations:} Normal font $x$, boldface lowercase font $\xb$, and boldface uppercase font $\Xb$ denote scalar, vector, matrix, respectively. $x_i$ (resp, $x_{i, j}$) shows element $i$ (resp.~$(i, j)$) of vector $\xb$ (resp. matrix $\Xb$).

\section{System Model}\label{sec:system}
This section describes the proposed AirCont system model and the differences with respect to the state-of-the-art scheme for the feedback control system over a wireless network.

\subsection{Control System Model}

\begin{figure}[t]
  \centering
  \includegraphics[width = 0.8\columnwidth]{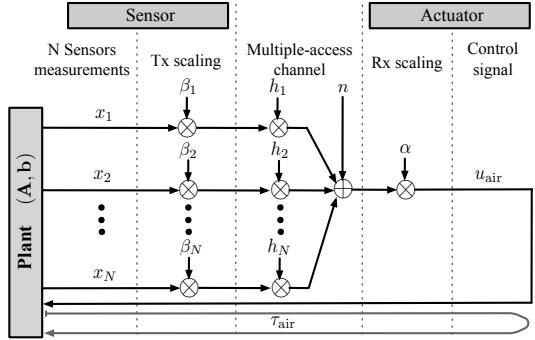}
   \caption{Over-the-air controller for closed-loop wireless control systems.}
   \label{fig:structure}
\end{figure}

We consider the problem of stabilizing a closed-loop control system composed of $N$ wireless sensors communicating with a single actuator where each sensor measures the state of a plant as depicted in Fig.~\ref{fig:structure}. The number of plant states is equivalent to the number of sensors, $N$.

We assume a standard continuous-time linear plant model
\begin{align}
\dot{\xb}(t) & = \Ab \xb(t) + \bb u(t)
\end{align}
with plant state $\xb \in \Rds^{N}$, state matrix $\Ab \in \Rds^{N \times N}$, input matrix $\bb \in \Rds^{N}$, and control signal $u \in \Rds$ applied to the plant. Each sensor attached to the plant periodically samples the plant's state and sends it by a sampling time of $\delta$.


The standard feedback controller receives the sampled plant state $\xb(t)$ after they are transmitted by the sensors and computes the discrete-time control input $u(k \delta) = -\kb^{\top} \xb(k \delta)$ where control gain $\kb \in \Rds^{N}$ if there is no communication delay. Due to the wireless communication delay $\tau$, which we assume to be shorter than the sampling period $\tau \leq \delta$, two control signals $u((k-1) \delta)$ and $u(k \delta)$ apply during $k$-th sampling period~\cite{Lunze14}. The control feedback signal is 
\begin{align}
u(t^{+}) & = -\kb^{\top} \xb(t - \tau), t \in \left\{ k \delta + \tau, k = 0, 1, 2, \ldots \right\} 
\end{align}
where $u(t^{+})$ is a piecewise continuous control signal and changes only at $k \delta + \tau$. We remark that the key interactive parameter between the network and the plant is the control signal, a linear combination of the sensors' signals.

\subsection{Proposed Control Signal Computation}
In AirCont, all sensors simultaneously transmit their data so that the actuator receives the control signal directly over the wireless channel based on the over-the-air computation concept. The scheme works as follows: each sensor $i$ scales the measured signal  $x_{i} \in \Rds$, representing the plant state, by the Tx scaling factor $\beta_i \in \Rds$ and concurrently transmits it to the actuator through a wireless channel with the channel coefficient $h_{i} \in \Rds$, as illustrated in Fig.~\ref{fig:structure}. The receiver adds additive white Gaussian noise (AWGN) $n \in N(0, \sigma^2)$. The actuator then scales the received signal with the Rx scaling factor $\alpha$ to compute the weighted sum of sensor data as
\begin{align}
u_{\rm air}(t^{+}) = - \alpha \left( (\hb  \odot \betab)^{\top}  \xb(t - \tau_{\rm air}) + n \right) \label{eq:u_oac} 
\end{align}
where $\tau_{\rm air}$ is the network delay, $n \in N(0, \sigma^2)$ is the receiver's AWGN and $\odot$ denotes Hadamard product between the Tx scaling factor and the channel coefficient for each sensor to the actuator. The Rx scaling factor $\alpha$ applies to both the signal and the noise terms. We consider a maximum transmit power constraint of each sensor, $\beta_i^2 \leq \overline{p}$, so, $\beta_i \in [0, \sqrt{\overline{p}}]$. Furthermore, we make the natural assumption that the sensors know the channel coefficients, and their transmissions are synchronized. The network delay of AirCont is only one single time slot $\tau_{\rm air} = T_s$ where $T_s$ denotes the duration of the time slot, since it leverages simultaneous coherent transmission from multiple sensors to the actuator.

As a benchmark, we consider a state-of-the-art scheme where each sensor attached to the physical plant sends the scaled signal $\beta_i x_{i}$ separately to the controller. Through the time slot allocated by a static scheduler, each sensor is only allowed to transmit once within a time frame cycle $\delta$, which is equivalent to the sampling period. To compute the control signal, the controller scales the received signal of sensor $i$ with the Rx scaling factor $\alpha_{s,i}$. The wireless channel between sensor $i$ and controller has a channel coefficient $h_{i}$ with AWGN $n_{s, i} \in N(0, \sigma_{s,i}^2)$. The controller then transmits the computed control signal to the actuator where we consider the channel coefficient $h_a$ with AWGN $n_{a} \in N(0, \sigma_{a}^2)$ for the controller-to-actuator link. Eventually, the actuator scales the received control signal with $\alpha_a$ to compensate the channel $h_a$. Hence, the resulting control signal of the state-of-the-art scheme is 
\begin{align}
u_{\rm sota}(t^{+}) = -  \alpha_a \left( h_a \alphab_s^{\top} \Db \xb(t - \tau_{\rm sota}) + \nb_s \right) + n_a  \label{eq:u_tra}
\end{align}
where $\Db$ is a diagonal matrix with $d_{i, i} = h_i \beta_i$ and $\tau_{\rm sota}$ is the network delay for the state-of-the-art scheme.

There are two main sources of network delay for the state-of-the-art scheme, namely the sensor-to-controller delay $\tau_{\rm sc}$ and the controller-to-actuator delay $\tau_{\rm ca}$. The end-to-end delay between sampling instance and actuating instance is the sum of the sensor-to-controller delay and controller-to-actuator delay, $\tau_{\rm sota} = \tau_{\rm sc} + \tau_{\rm ca}$. Since the state-of-the-art scheme requires a transmission from all sensors to the controller then to the actuator, the minimum delay in actuating the control signal in each sampling instance is $\tau_{\rm sota} = (N + 1) T_s$. In contrast, AirCont integrates the communication and computation of a large number of sensor data in one time slot. It does not rely on the dedicated controller since the actuator directly adapts the control signal as a linear combination of the sensor data. Hence, it reduces 2-hop communication, namely, sensor-to-controller and controller-to-actuator, to a single hop, namely, sensor-to-actuator.


\section{Performance Analysis}\label{sec:analysis}
In this section, we first derive the achievable stability region restricted by the operational constraints of AirCont and the state-of-the-art scheme.  We then investigate the computation error of the control signal using different scaling control policies. 

\subsection{Stability Analysis}\label{sec:analsta_fea}
A higher sampling rate is generally desirable in the discrete-time system since it approximates well the continuous-time system. However, a higher sampling rate increases the network delay due to the possible congestion~\cite{Park18}. The WNCS design needs to find a sampling rate and delay that can both ensure the control stability and be achievable by the wireless network.

\Blue{Quantifying the stability boundary of the control system with respect to the sampling period $\delta$ and delay $\tau$ is useful to understand the control performance tradeoffs. We define two distinct notions, namely, \textit{maximum stability region} and \textit{achievable stability region}.} 

\begin{definitionPan}[Maximum Stability Region]\label{def: msr}
\Blue{It is the set of values of sampling period $\delta$ and delay $\tau$, which guarantees the control stability of the closed-loop system.}
\end{definitionPan}

\begin{definitionPan}[Achievable Stability Region]\label{def: asr}
\Blue{It is the set of values of sampling period $\delta$ and delay $\tau$ that can be supported by the wireless network, which guarantees the control stability of the closed-loop system.}
\end{definitionPan}

We extend the approach in~\cite{Lunze14} to analyze the stability of the AirCont-based system and the benchmark state-of-the-art scheme. By considering the sampled system with sampling period $\delta$, the expected linear difference equation becomes 
\begin{align}
\overline{\xb}((k+1)\delta) = \Phi \overline{\xb}(k \delta) + \Gamma_0(\tau) \overline{u}(k \delta) + \Gamma_1 (\tau) \overline{u}((k-1)\delta) \nonumber
\end{align}
where $\overline{\xb}$ and $\overline{u}$ are the expected value of $\xb$ and $u$ with respect to noise factors,  $\Phi = \er^{\Ab \delta},$
\begin{align}
\Gamma_0(\tau)   = \int_{0}^{\delta - \tau} \er^{\Ab s} \, \Bb \, \dr s \,,  \,\,\,\,  \Gamma_1(\tau)  = \int_{\delta - \tau}^{\delta} \er^{\Ab s} \, \Bb \, \dr s  \,. \nonumber
\end{align}
$\Gamma_0(\tau)$ and $\Gamma_1(\tau)$ are the input matrix of two control signals $\overline{u}(k \delta)$ and $\overline{u}((k-1)\delta)$ due to the network delay $\tau$, respectively. Remind that the end-to-end network delay from the sampling instance is shorter than the sampling period, $\tau \leq \delta$.

By defining $\zb (k \delta) = [\overline{\xb}^{\top}(k \delta)  \; \overline{u}((k-1) \delta)]^{\top}$ as the augmented state vector, the augmented system becomes
\begin{align}
\zb ((k+1) \delta) = \tilde{\Phi} \zb(k \delta)\,. \label{eq:zb}
\end{align}
The discrete-time linear system is asymptotically stable (in fact, exponentially stable) if all eigenvalues of $\tilde{\Phi}$ have norm strictly less than one, i.e., the spectral radius $\rho(\tilde{\Phi}) < 1$. 

By using the AirCont scheme in Eq.~\eqref{eq:u_oac}, the augmented system matrix in~\eqref{eq:zb} becomes 
\begin{align}
\tilde{\Phi}_{\rm air}  = 
\begin{bmatrix}
\Phi - \Gamma_0(\tau_{\rm air}) \alpha  (\hb  \odot \betab)^{\top}    & \Gamma_1(\tau_{\rm air})  \\
- \alpha  (\hb  \odot \betab)^{\top}  & 0 
\end{bmatrix} \,. \label{eq:ext_mat_oac}
\end{align}
The matrix $\tilde{\Phi}_{\rm air}$ depend on both control aspects (continuous-time plant dynamics $(\Ab, \bb)$ and sampling period $\delta$) and wireless communication aspects (channel coefficient $\hb$, delay $\tau_{\rm air}$ , and Tx-Rx scaling factors $(\betab, \alpha)$).

In a similar way, the stability condition of the state-of-the-art scheme is analyzed by considering Eq.~\eqref{eq:u_tra}, which gives the augmented system matrix
\begin{align}
\tilde{\Phi}_{\rm sota}  = 
\begin{bmatrix}
\Phi - \Gamma_0(\tau_{\rm sota}) \alpha_a h_a \alphab_s^{\top} \Db  & \Gamma_1(\tau_{\rm sota})  \\
- \alpha_a h_a \alphab_s^{\top} \Db  & 0 
\end{bmatrix} \,. \label{eq:ext_mat_conv}
\end{align}


\Blue{The maximum stability region is determined by the spectral radius of the extended closed-loop system matrices of Eqs.~\eqref{eq:ext_mat_oac} and~\eqref{eq:ext_mat_conv}, whereas the achievable stability region is a subset of the maximum stability region restricted to the feasibility constraints $\tau_{\rm air} \leq \delta$ and $\tau_{\rm sota} \leq \delta$.}

\subsection{Mean Square Error Analysis}\label{sec:mse}
This section investigates the computation error (measured by the Mean Square Error (MSE)) of different Tx-Rx scaling policies subject to individual Tx scaling constraints of sensors. Some existing works~\cite{Abari16, Liu20} have already considered the Tx-Rx scaling optimization problem for the over-the-air computation, where the objective is to minimize the computation error subject to the Tx scaling limits of sensors. Inspired by~\cite{Abari16, Liu20}, we propose a minimization of the average MSE with respect to the control signal $\kb^{\top} \xb$. To simplify the analysis, we assume that measured signals $x_i$ are independent and follow the standard normal distribution $N(0, 1)$. 

We first analyze the computation distortion of the control signal using AirCont. The MSE between  Eq.~\eqref{eq:u_oac} and the control signal  $\kb^{\top} \xb$ is 
\begin{align}
\MSE_{\rm air} & = \Eds \left[ \left\vert \alpha \left( (\hb  \odot \betab)^{\top}  \xb + n \right)  - \kb^{\top} \xb \right\vert^2 \right]  \nonumber \\
& =  \left(\alpha (\hb  \odot \betab)  - \kb \right)^{\top} \left(\alpha (\hb  \odot \betab)  - \kb \right) + \sigma^2 \alpha^2\label{eq:oac_mse}
\end{align}
where the expectation of MSE is calculated with respect to the distributions of $\xb$ and $n$. We obtain the optimal Tx-Rx scaling policy of the MSE minimization without any constraints on the Tx scaling factor. The optimal Tx-Rx scaling factors are $\beta_i \rightarrow  k_i/(\alpha h_i)$ and $\alpha \rightarrow 0$ since these scaling factors minimize the first and second terms of Eq.~\eqref{eq:oac_mse}, respectively. By putting the optimal solutions of Tx-Rx scaling factors to Eq.~\eqref{eq:oac_mse}, the MSE using AirCont approaches $0$. However, the optimization problem becomes non-convex with the Tx scaling limits. We extend the existing Tx-Rx scaling policy~\cite{Liu20} to the MSE minimization problem of the weighted sum of sensor data and evaluate its performance.

As a benchmark, we investigate the optimal scaling factors of Tx scaling factor $\betab$ of sensors, Rx scaling factor $\alphab_s$ of controller, and Rx scaling factor $\alpha_a$ of actuator for the state-of-the-art scheme. By considering Eq.~\eqref{eq:u_tra}, the computation distortion, MSE, of the control signal $\kb^{\top} \xb$ is
\begin{align}
\MSE_{\rm sota} = &\, \Eds \left[ \left\vert \alpha_a \left( h_a \alphab_s^{\top}  \left( \Db  \xb + \nb_s \right) + n_a \right)- \kb^{\top} \xb \right\vert^2 \right]  \nonumber \\
= &  \, (\alpha_a h_a \Db \alphab_{s} - \kb)^{\top} (\alpha_a h_a \Db \alphab_{s} - \kb) \nonumber \\ 
& + \alpha_a^2 h_a^2 \sigma_s^2 \alphab_s^{\top} \alphab_s + \alpha_a^2 \sigma_a^2 \label{eq:tra_mse}
\end{align}
where the expectation of MSE is calculated with respect to $\xb$, $\nb_s$, and $n_a$. Recall that the diagonal matrix $\Db$ depends on the channel coefficient $\hb$ and the Tx scaling factor $\betab$ as $d_{i, i} = h_i \beta_i$. The MSE of the state-of-the-art scheme converges to $0$ when $\beta_{i} \rightarrow k_i/(\alpha_a \alpha_{s, i} h_i), \alpha_a \rightarrow 0$ and $\alpha_{s, i} \rightarrow 0$ if the Tx scaling factor is not constrained.

We separate the constrained optimization problem into two sub-problems, namely, $\betab^{\ast}$ and $\alphab_{s}^{\ast}$ for sensor-to-controller and $\alpha_a^{\ast}$ for controller-to-actuator. By considering Eq.~\eqref{eq:tra_mse}, $\hb, \kb$ and $\sigma_s^2$, we reformulate the optimization problem to optimize $\betab$ and $\alphab_s$ for sensor-to-controller communications where the objective function is 
\begin{align}
\MSE_{\rm sota}^{\rm sc} & = (\Db \alphab_{s} - \kb)^{\top} (\Db \alphab_{s} - \kb) + \sigma_s^2 \alphab_s^{\top} \alphab_s \label{eq:tra_mse_sc}
\end{align}
subject to the Tx scaling limits. To minimize the first term of Eq.~\eqref{eq:tra_mse_sc}, we optimize each pair of $\beta_{i}$ and $\alpha_{s, i}$ from sensor $i$ to the controller to meet $\Db \alphab_s = \kb$. Since the noise term of Eq.~\eqref{eq:tra_mse_sc} only depends on $\alphab_s$, the optimal values are $\beta_{i}^{\ast} = \sqrt{\overline{p}}$ and $\alpha_{s, i}^{\ast} = (h_i k_i \beta_{i}^{\ast})/((h_i\beta_{i}^{\ast})^2 + \sigma_s^2)$. Substituting $\beta_{i}^{\ast}$ and $\alpha_{s, i}^{\ast}$ to Eq.~\eqref{eq:tra_mse}, we obtain the optimal scaling factor of the actuator as $\alpha_a^{\ast} = (h_a \alphab_s^{{\ast}\top} \Db^{\ast} \kb) / (h_a^2 \alphab_s^{{\ast}\top} (\Db^{\ast})^2 \alphab_s  + h_a^2 \sigma_s^2 \alphab_s^{{\ast}\top} \alphab_s^{\ast} + \sigma_a^2)$ where the diagonal matrix $\Db^{\ast} $ is calculated using $\betab^{\ast}$.

We note that the sensor-to-controller and controller-to-actuator scheme significantly increases the complexity of the operation compared to AirCont since the actuator requires the knowledge of the sensor-to-controller communication including $\hb, \betab^{\ast}$, $\alphab_s^{\ast}$ and $\sigma_s^2$ in addition to $h_a, \sigma_a^2$ and $\kb$.


\section{Performance Evaluation}\label{sec:eval}
This section evaluates the stability region and the average MSE of AirCont compared to the state-of-the-art scheme.

\begin{figure}[t]
  \centering
  \subfigure[Stability region using AirCont.]
  {
  \psfrag{x}[][]{\footnotesize{$\delta$ (s)}}
  \psfrag{y}[][]{\footnotesize{$\tau/\delta$}}      
  \includegraphics[width = 0.43\columnwidth]{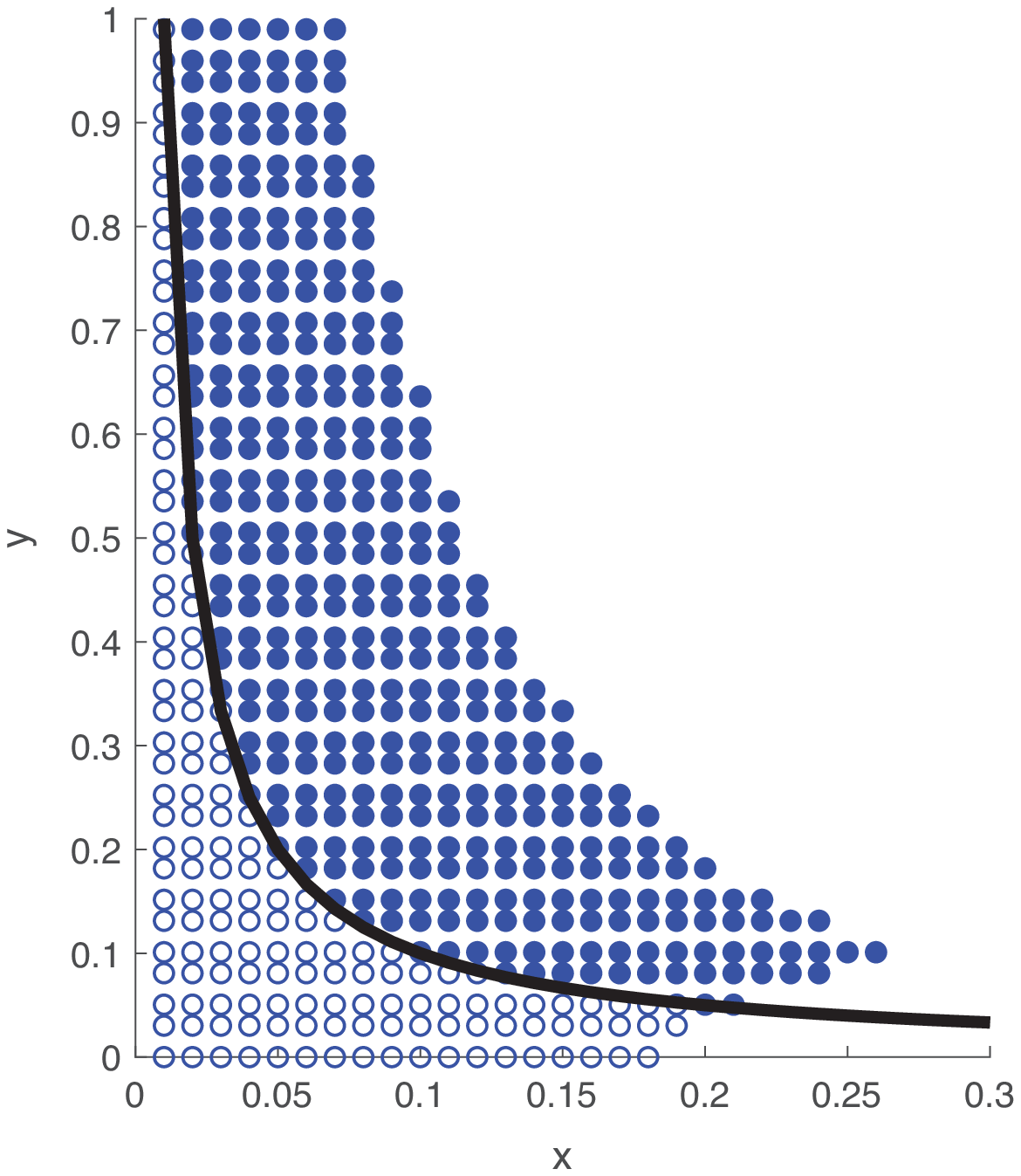}
  \label{fig:sta_oac}
  }
  \subfigure[Stability region using state-of-the-art scheme.]
  {
  \psfrag{a}[][]{\footnotesize{$\delta$ (s)}}
  \psfrag{b}[][]{\footnotesize{$\tau/\delta$}}      
  \includegraphics[width = 0.43\columnwidth]{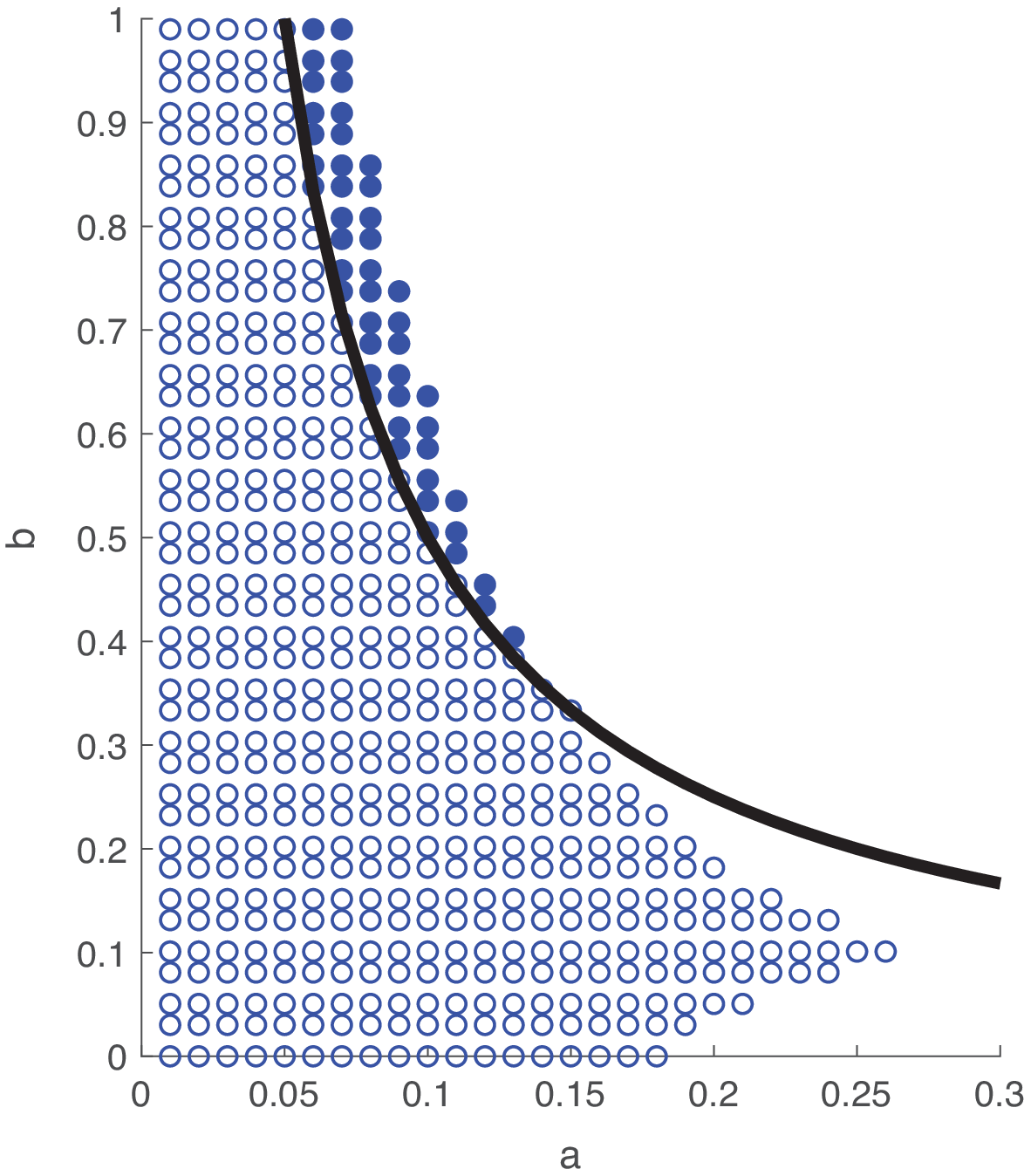}  
  \label{fig:sta_con}
  }
  \caption{\Blue{Maximum stability region and achievable stability region using AirCont and the state-of-the-art scheme with different sampling period $\delta$ and delay $\tau$. All circular markers present the maximum stability region for a given sampling period and delay. To quantify the network effects, we represent the achievable stability region of WNCS using AirCont and the state-of-the-art scheme with filled circular markers. The solid line shows the lower bound of the feasible sampling period and delay $\tau \leq \delta$ using each of the schemes.} } \label{fig:stability}
\end{figure}



\subsection{Stability Performance}
\Blue{We first quantify the maximum stability region and achievable stability region of the control system with respect to the sampling period $\delta$ and delay $\tau$. We consider the position control for the ball and beam system where the linear time-invariant model with $N=4$ sensors and $1$ actuator is used~\cite{Goodwin00}. Fig.~\ref{fig:stability} shows the maximum stability region and the achievable stability region of both AirCont and state-of-the-art scheme with different sampling period $\delta$ and delay $\tau$ for $\tau \leq \delta$. For a fair comparison, the control gains for AirCont and state-of-the-art scheme are set to be equal, with $\alpha (\hb  \odot \betab)^{\top} = \alpha_a h_a \alphab_s^{\top}  \Db = [6.67, 11.09,  41.15, 11.27]$ where $\Db$ is a function of $\hb$ and $\betab$ without any noise.}

\Blue{Given sampling period $\delta$ and delay $\tau$, we plot the stability region by evaluating the spectral radius of extended closed-loop system matrices of Eqs.~\eqref{eq:ext_mat_oac} and~\eqref{eq:ext_mat_conv}. In Fig.~\ref{fig:stability}, a point is marked with circular markers to represent the maximum stability region. As expected, the maximum stability regions of AirCont and state-of-the-art scheme are identical since the network effects are not explicitly considered. To quantify the network effects,  the achievable stability regions of both AirCont and state-of-the-art scheme are reported with filled circular markers. The solid line shows the lower bound of the feasible sampling period and delay $\tau \leq \delta$ using AirCont and the state-of-the-art scheme in Fig.~\ref{fig:stability}.}

\Blue{In Fig.~\ref{fig:stability}, a lower sampling period and lower delay are generally desirable for the maximum stability region since the control system ensures the stability for a delay up to the full sampling period for $\delta \leq 0.07$s. As the sampling period $\delta$ increases, the upper bound of $\tau/\delta$ of the maximum stability region is considerably reduced. Note that the control system becomes unstable even without delay for $\delta > 0.26$s. However, it is not trivial to quantify the maximum stability boundaries as we observe that the lower delay $\tau/\delta \leq 0.11$ is worse for the control stability for $0.18 \mathrm{s} \leq \delta \leq 0.26$s, as shown in Fig.~\ref{fig:stability}.}

\Blue{The network performance heavily affects the achievable stability region of the control system. The lower delay is not achievable for the short sampling period due to the fundamental congestion of the network performance. In Fig.~\ref{fig:sta_con}, the state-of-the-art scheme is not able to provide the control stability of the plant for $\delta < 0.05$s due to the minimum network delay constraint $\tau_{\rm sota} = 0.05 \sr \leq \delta$. \Blue{On the other hand, the minimum sampling period of AirCont is $\delta = 0.01$s, significantly lower than the one of the state-of-the-art scheme in Fig.~\ref{fig:sta_oac}.}}

\Blue{A large achievable stability region improves operating robustness against uncertain losses and delays, and energy efficiency by reducing the sampling rate. By comparing Figs.~\ref{fig:sta_oac} and~\ref{fig:sta_con}, the achievable stability region using AirCont is $7.3$ times larger than that using the state-of-the-art scheme. The main reason is that AirCont only takes a single time slot to compute and communicate the control signal for the sensor-actuator, independently of the number of sensors.}




\begin{figure}[t]
  \centering
  \subfigure[Average control MSE versus different Tx scaling limits.]
  {
  \psfrag{x}[][]{\footnotesize{Peak power limit, $\overline{p}$}}
  \psfrag{y}[][]{\footnotesize{Average control MSE}}      
  \includegraphics[width = 0.83\columnwidth]{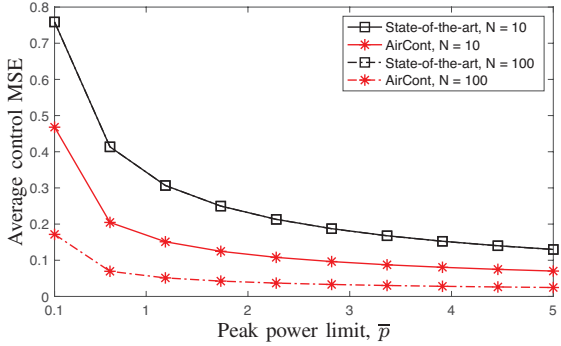}  
  \label{fig:mse_power}
  }
  \subfigure[Average control MSE versus different noise variance.]
  {
  \psfrag{x}[][]{\footnotesize{Noise variance, $\sigma^2 = \sigma_s^2 = \sigma_a^2$}}
  \psfrag{y}[][]{\footnotesize{Average control MSE}}        
  \includegraphics[width = 0.83\columnwidth]{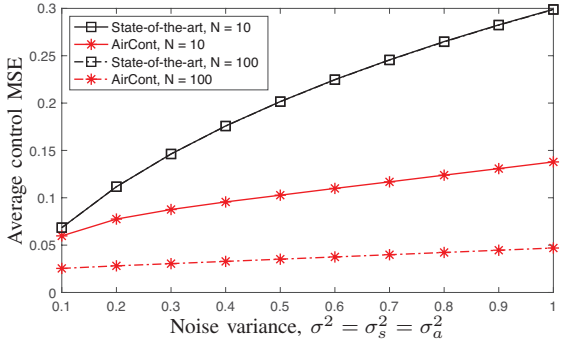}
  \label{fig:mse_var}
  }
\caption{\Blue{Average control MSE of both AirCont and state-of-the-art scheme with $N = 10, 100$ for various Tx scaling limits and noise variance.}} \label{fig:mse}
\end{figure}

\subsection{MSE Performance}
Now, we investigate how AirCont and the state-of-the-art scheme behave with different Tx scaling limits, noisy channels, and a number of sensors on the average MSE. We set the peak transmit power $\overline{p} = 2.5$ and the noise variance $\sigma^2 = \sigma_s^2 = \sigma_a^2 = 0.5$ unless otherwise stated. The channel coefficient is i.i.d Rayleigh fading with unit variance. We set the control gain $\kb$ with a uniformly generated random value between $0$ and $100$. To normalize the effect of the control gain, we define the average control MSE as the expected value of MSE dividing with the sum of the squared control gain, namely, $\Eds \left[ \MSE \right] /(\kb^{\top} \kb)$. We calculate the average control MSE using extensive Monte Carlo simulations with random channel and control gain realizations.

\Blue{Fig.~\ref{fig:mse_power} depicts the average control MSE of both AirCont and state-of-the-art scheme with number of sensors $N = 10, 100$ while varying the peak power limits $\overline{p} = 0.1, \ldots, 5$. The AirCont scheme outperforms consistently well the state-of-the-art scheme with respect to varying Tx scaling constraints. While the average control MSEs of both schemes decrease approximately exponentially with the peak power limits, the gap between them increases as the power constraint becomes strict.} The AirCont scheme has control MSE approaching $0$ as the peak power limits relaxes and the benefits of AirCont drastically improve as the number of sensors increases, while this effect is negligible for the state-of-the-art scheme.

Fig.~\ref{fig:mse_var} plots the average control MSEs of both AirCont and state-of-the-art scheme with number of nodes $N = 10, 100$ while varying noise variances  $\sigma^2 = \sigma_s^2 = \sigma_a^2 = 0.1, \ldots, 1$. \Blue{The average control MSEs of both schemes roughly increases linearly with the noise variance.} The AirCont scheme provides a remarkably lower average control MSE than that of the state-of-the-art scheme throughout the whole considered range.

Comparing the MSE slopes between AirCont and state-of-the-art scheme, the average control MSE of AirCont is less sensitive to the noise variance than the state-of-the-art scheme. The state-of-the-art scheme possibly emphasizes the noise effect of the multi-hop communication for sensor-to-controller-to-actuator links due to the channel distortion and the Tx scaling limit, as discussed in Section~\ref{sec:mse}. The MSE gain of AirCont increases as the noise variance increases. Furthermore, while AirCont considerably improves the average control MSE for a large number of nodes $N=100$, this effect is negligible for the state-of-the-art scheme due to the presence of uncorrelated noise for each sensor transmission. \Blue{By analyzing Fig.~\ref{fig:mse}, the AirCont scheme is particularly attractive for WNCS relying on dense sensor networks with strict Tx scaling limits and noisy channels.}


\begin{figure}[t]
  \centering
  \psfrag{x}[][]{\footnotesize{Time (s)}}
  \psfrag{y}[][]{\footnotesize{Response}}          
  \includegraphics[width = 0.88\columnwidth]{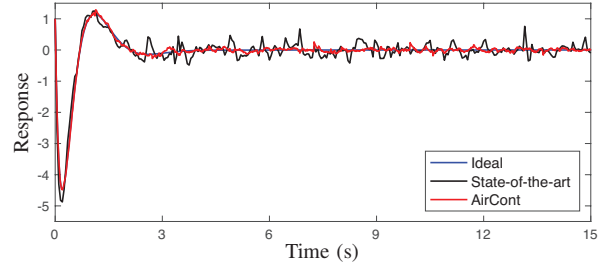}
   \caption{Plant outputs of the ball and beam control system using an ideal solution, state-of-the-art scheme, and AirCont}    
   \label{fig:response}
\end{figure}

Finally, we illustrate how AirCont improves the control performance as a case study using the ball and beam system~\cite{Goodwin00}. In Fig.~\ref{fig:response}, we show $4$-th plant output of the control system using an ideal solution, state-of-the-art scheme, and AirCont. Note that the ideal solution means no delay and no channel noise of the communication. The plant output of AirCont is very similar to that of the ideal solution. The oscillations of the state-of-the-art scheme imply that it does not essentially guarantee good control performance. The network operating region of the state-of-the-art scheme is closer to the unstable region, as shown in Fig.~\ref{fig:sta_con} due to the minimum sampling period $\delta = 0.05$s. \Blue{Furthermore, the control stability using the state-of-the-art scheme is significantly vulnerable to the noise since it increases the computation error of the control signal.}

\section{Conclusion}\label{sec:conc}
This paper presents AirCont, a novel paradigm that eliminates the control unit and computes the control signal for closed-loop wireless control systems, adopting the over-the-air computation concept. As opposed to the sensor-to-controller and controller-to-actuator communication of state-of-the-art schemes, AirCont effectively integrates communication and computation by harnessing interference for computing the control signal as the weighted sum of the sensor data. This approach simplifies the control system operation, design, and analysis using direct sensor-to-actuator communications without relying on a dedicated control unit. We demonstrated that AirCont can dramatically improve the achievable stability region compared to the state-of-the-art scheme. Moreover, numerical results confirmed that AirCont substantially reduces the computation error of the control signal for various power limits, channel conditions, and number of nodes.
  
Previous works on over-the-air function computation focus on optimizing the Tx-Rx scaling factors to minimize the computation error. However, this solution does not guarantee the optimal control cost. Inspired by this observation, we aim at developing a Tx-Rx scaling policy to provide robust control performance over uncertain channel distortion.


\bibliographystyle{IEEEtran}
\bibliography{ref_oac}

\end{document}